\documentclass{article}
\usepackage{graphicx} 
\usepackage{amsmath}
\usepackage{bbm}

\title{Numerical complexity of helix unraveling algorithm for charged particle tracking}
\author{Kacper Topolnicki and Tomasz Bold} 
\date{May 2023}

\begin{document}

\maketitle

\begin{abstract}
This paper describes a procedure for a realistic estimation of the number of iterations
in the main loop of a recent particle detection algorithm from \cite{Topolnicki_2022}.
The calculations are based on a Monte Carlo simulation of the 
ATLAS inner detector. The resulting estimates of numerical complexity 
suggest that using the procedure from \cite{Topolnicki_2022} for online 
triggering is not feasible. There are however some areas, such as triggering for 
particles in a specific sub-domain of the phase space, where using this
procedure might be beneficial.
\end{abstract}

\section{Introduction}
\label{introduction}

The fast and accurate recognition of helical charged particle
tracks in data collected by modern colliders such as the
High Luminosity LHC \cite{EPJ_WOC_2018} is a crucial step
in uncovering physics that is potentially beond the Standard model.
Before starting athe start of a new experiment, members of the ATLAS Collaboration
prepare new methods for handling the large amounts of data collected by the
detector in real time and perform tracking
\cite{Collaboration_2008,Collaboration2_2017,Collaboration1_2017,ATLASTDAQAmendment}.
Particles with long lifetimes, that
decay at large distances from the beam line,
are of particular interest \cite{Bobrovskyi2013}.
A review of methods used for particle tracking is available in \cite{newbook}.
The algorithm from \cite{Topolnicki_2022} was proposed as a novel
way to search for such charged particle tracks in data from
high energy physics detectors submerged in a uniform magnetic field. 
It is designed to be agnostic to the origin of particle tracks
making it possible to detect particles with longer lifetimes.

The algorithm can be divided into independent iterations. 
Each individual iteration searches the data for helical charged
particle tracks with a given set of parameters and
consists of three steps:
\begin{enumerate}
    \item Load the input data in the form of Cartesian coordinates of detected track positions: {$\mathbbm{D} = \{(x_{i} , y_{i} , z_{i}) , i = 1 \ldots N\}$}.
    \item Calculate the image space {$\mathbbm{D}' = \{u_{x_{c} , y_{c} , \nu}(x_{i} , y_{i} , z_{i}) , i = 1 \ldots N\}$} by mapping
    a special function $u_{x_{c} , y_{c} , \nu}$ over $\mathbbm{D}$. This function
    has an additional dependence on three parameters of the helix (these are discussed in more detail below): the position of the helix axis $x_{c}$, $y_{c}$ and the helix pitch $\nu$. If a helical charged particle track which matches the additional parameters
    of $u_{x_{c} , y_{c} , \nu}$ is present in the data collected by the detector
    then these points will be mapped into a straight line along $\hat{z}$.
    \item A peak detected on a $\hat{x} - \hat{y}$ histogram of $\mathbbm{D}'$ indicates the existence of a helical particle
    track in $\mathbbm{D}$ with parameters $x_{c} , y_{c} , \nu$.
\end{enumerate}
In these steps $N$ is the total number of Cartesian points from the detector and $u_{x_{c} , y_{c} , \nu}$
is a special transformation that takes a helix with given parameters $x_{c} , y_{c} , \nu$ and turns it
into a straight line along $\hat{z}$ making it detectable as a histogram peak.

\begin{figure}
    \centering
    \includegraphics[width=0.65 \textwidth]{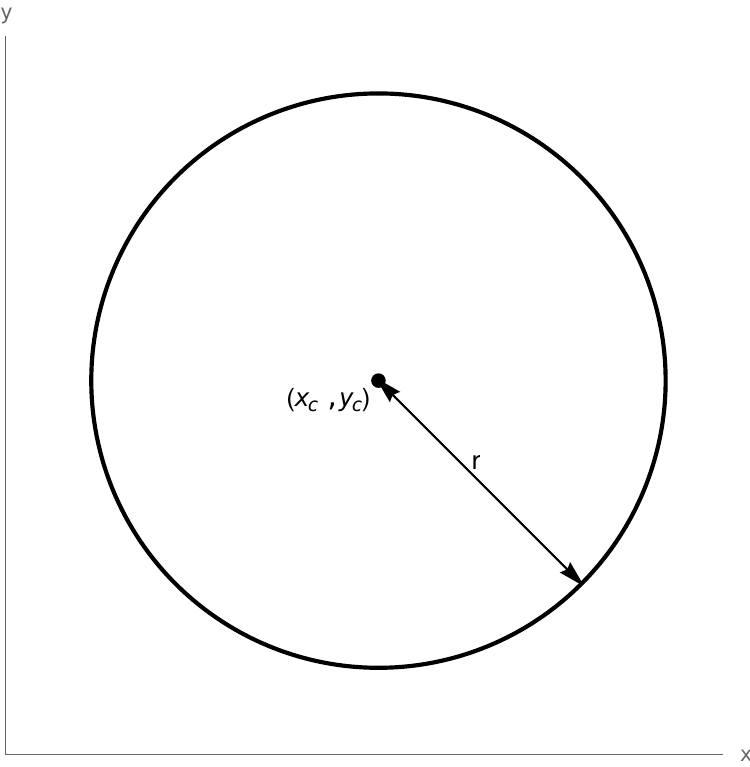}
    \caption{
    A helical particle track projected on the $\hat{x} - \hat{y}$ plane is a circle or a fragment of a circle. The detector is centered around
    the origin and the beam line is perpendicular to the diagram. 
    The helix axis position is $(x_{c} , y_{c})$. The third parameter $\nu$ is not illustrated and
    plays the role of the helix pitch. This can be seen in the explicit form of a point on the helix:
    $\left(x_{c} , y_{c} , 0\right) + \left(r \cos{\left(-\frac{\nu (z - z_{0})}{r}\right)} , r \sin{\left(-\frac{\nu (z - z_{0})}{r}\right)} , z\right)$ where $r$ is the helix radius and $z_{0}$ fixes the helixes position along $\hat{z}$. 
    }
    \label{fig:my_label}
\end{figure}

The three parameters of helical tracks used in the procedure are illustrated and described on Figure \ref{fig:my_label}.
The explicit form of the "unraveling" function $u$ was given in \cite{Topolnicki_2022} as:
\begin{align} \nonumber
	u_{x_{c} , y_{c} , \nu}(x , y , z) &:= 
	(x_{c} , y_{c} , 0) 
	+ R_{\hat{z}}\left(\frac{z \nu}{\sqrt{(x - x_{c})^{2} + (y - y_{c})^{2}}}\right) \\ 
	& \left( (x , y , z) - (x_{c} , y_{c} , 0) \right)
	\label{u_transformation}
\end{align}
and is essentially a simple rotation $R_{\hat{z}}(\alpha)$ of a Cartesian point along $\hat{z}$ with a center of rotation at $(x_{c} , y_{c})$ in the $\hat{x} - \hat{y}$ plane. What makes this transformation useful is a careful
choice of the angle of rotation $\alpha$. This angle depends on the $\hat{z}$
coordinate and allows the detected Cartesian points from a particle track with parameters $x_{c}, y_{c}, \nu$ to be "unraveled" into a straight line along $\hat{z}$. This collection of points can be detected as a peak on a $\hat{x} - \hat{y}$ histogram. In this paper a slightly more general form of (\ref{u_transformation}) is used:
\begin{align} \nonumber
	u_{x_{c} , y_{c} , \nu}(x , y , z) &:= 
	(x_{c} , y_{c} , 0) 
	+ R_{\hat{z}}\left(\frac{(z - \bar{z}) \nu}{\sqrt{(x - x_{c})^{2} + (y - y_{c})^{2}}}\right) \\ 
	& \left( (x , y , z) - (x_{c} , y_{c} , 0) \right)
	\label{u_transformation1}
\end{align}
where the additional parameter $\bar{z}$ gives the flexibility to chose the fixed point of the transformation:
\begin{equation}
  u_{x_{c} , y_{c} , \nu}(x , y , \bar{z}) = (x , y , \bar{z}).      
\end{equation}
Setting $\bar{z} = 0$ turns (\ref{u_transformation1}) into (\ref{u_transformation}) and results in the fixed point being on the $z = 0$
plane. This can be problematic since this plane contains, in the Monte Carlo simulations used, the interaction point
and may result in
many peaks, close together on the $\hat{x} - \hat{y}$ histogram of $\mathbbm{D}'$ making them 
difficult to distinguish. More details about the algorithm can be found in \cite{Topolnicki_2022}. 

\section{Determining the step size}

As mentioned in Section \ref{introduction} the algorithm can be divided into independent iterations, 
each iteration searching the input data for tracks with a given set of parameters. In order to arrive
at a full implementation of the algorithm it is necessary to determine the change of
helix parameters from one iteration to another:
\begin{equation}
x_{c} , y_{c} , \nu \rightarrow x_{c}' , y_{c}' , \nu' .
\label{change}
\end{equation}
The original paper \cite{Topolnicki_2022} used an approximate approach based on dimensional 
analysis to determine the total number of these iterations. These results were not precise
and a new approach was needed. 

\begin{figure}
    \centering
    \includegraphics[width=0.65 \textwidth]{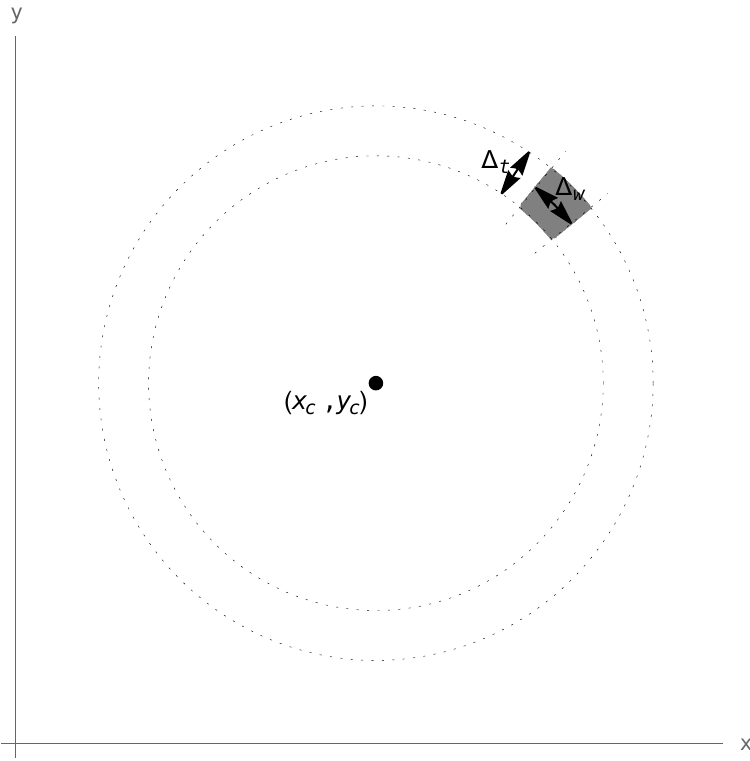}
    \caption{The shape of bins used to determine step sizes. A single bin is a $\Delta_{w}$ width fragment of a round slice centered around the helix axis $(x_{c} , y_{c})$ with thickness $\Delta{t}$.
    In the calculations used for this paper $\Delta_{w} = 10^{-4} \, m$ and $\Delta_{t} = 5 \times 10^{-5} \, m$.
    These numbers ensured that over TODO of helices were detected.} 
    \label{bn}
\end{figure}

\begin{figure}
    \centering
    \includegraphics[width=0.65 \textwidth]{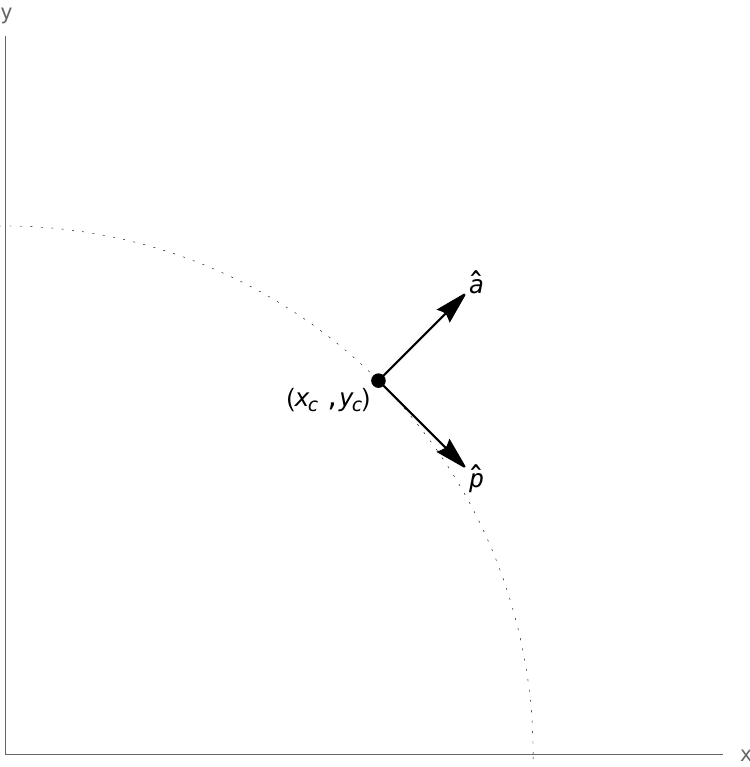}
    \caption{When creating a map of allowable parameter changes, the helix center
    $(x_{c} , y_{c})$ was moved in two directions: along the vector from the origin $\hat{a}$
    and perpendicular to this vector $\hat{p}$.}
    \label{ch}
\end{figure}

In this paper we use a more direct method
and base it on based on a realistic monte carlo simulation
of a detector, the Open Data Detector \cite{ODD}. The same simulation was also used in \cite{Topolnicki_2022}.
The following procedure was used to determine the allowable parameter step sizes:
\begin{enumerate}
\item Chose a reference trajectory from an random event generated by the simulation.
\item Set $\bar{z}$ to match one point on the reference trajectory. 
    This step will make it easy to calculate the $\hat{x} - \hat{y}$ position of the "unraveled"
    trajectory in $\mathbbm{D}'$.
\item Set the step sizes $\text{d}x_{c} , \text{d}y_{c} , \text{d}\nu = 0 , 0 , 0$.
\item \label{loop1} All parameters $x_{c} , y_{c} , \nu$ of the reference trajectory are known. Use 
    $x_{c} + \text{d}x_{c} , y_{c} + \text{d}y_{c}, \nu + \text{d}\nu$ to "unravel" the whole event.
    If the unraveling parameters don't match the reference trajectory parameters exactly, the reference
    helix will not unravel into a perfectly streight line.
\item Look for peaks in a bin centered at $(x_{c} , y_{c})$. Bin shapes and sizes are
    shown in Figure \ref{bn}.
\item Depending on if a peak is present or not, increase or decrease the step sizes
    $\text{d}x_{c} , \text{d}y_{c} , \text{d}\nu$ accordingly. 
    In practice the step sizes $\text{d}x_{c} , \text{d}y_{c} , \text{d}\nu$ are chosen to move the helix axis $(x_{c} , y_{c})$ 
    in two perpendicular directions
    as shown in Figure \ref{ch}.
\item Repeat from \ref{loop1} to determine the maximum change $\text{d}x_{c} , \text{d}y_{c} , \text{d}\nu$ in reference trajectory
    helix parameters $x_{c} , y_{c} , \nu$ for which the reference trajectory is still detected.
\end{enumerate}
The condition for the step size is that before and after (\ref{change})
the helix is still detectable.
Using this condition, the end result of the 7 step procedure above is a map of maximum allowable step sizes 
for different helix parameters $x_{c} , y_{c} , \nu$. 

\begin{figure}
    \centering
    \includegraphics[width=0.85 \textwidth]{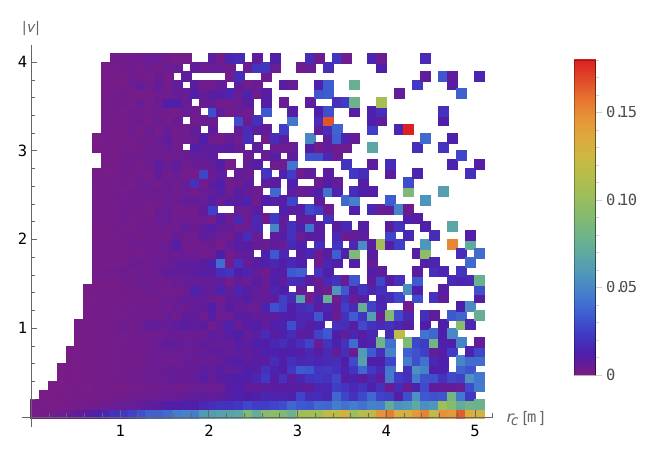}
    \caption{PRELIMINARY Map of allowable shifts $\text{d}a^{max}(r_{c} , |\nu|)$ , in meters, of the helix center $(x_{c} , y_{c})$ in the $\hat{a}$
    direction from Figure \ref{ch}. The horizontal axis $r_{c}$ is the distance of the heilx axis 
    from the origin. The vertical axis is the absolute value of the helix pitch $\nu$.} 
    \label{along}
\end{figure}

\begin{figure}
    \centering
    \includegraphics[width=0.85 \textwidth]{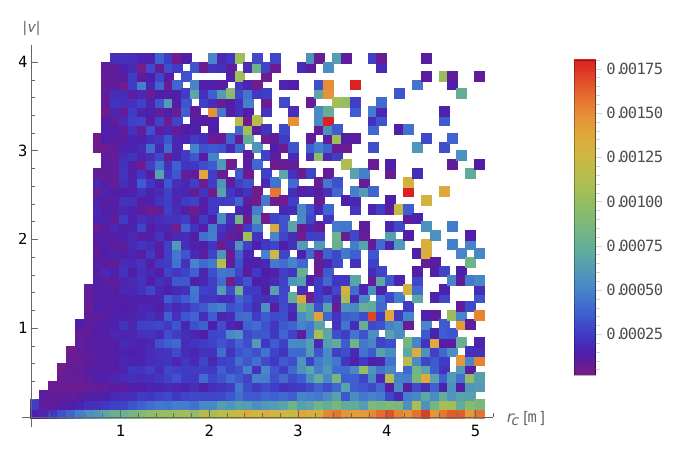}
    \caption{PRELIMINARY Map of allowable shift $\text{d}p^{max}(r_{c} , |\nu|)$, in meters, of the helix center $(x_{c} , y_{c})$ in the $\hat{p}$
    direction from Figure \ref{ch}. The horizontal axis $r_{c}$ is the distance of the heilx axis 
    from the origin. The vertical axis is the absolute value of the helix pitch $\nu$.} 
    \label{perp}
\end{figure}

\begin{figure}
    \centering
    \includegraphics[width=0.85 \textwidth]{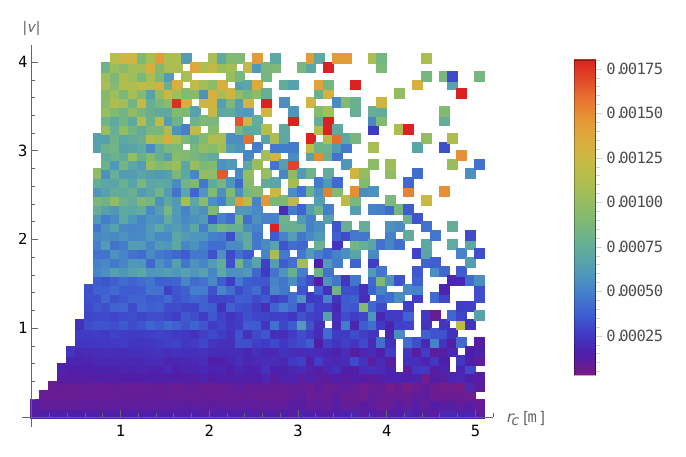}
    \caption{PRELIMINARY Map of allowable shifts, in meters, of the helix center $(x_{c} , y_{c})$ in the $\hat{p}$
    direction from Figure \ref{ch}. The horizontal axis $r_{c}$ is the distance of the heilx axis 
    from the origin. The vertical axis is the absolute value of the helix pitch $\nu$.} 
    \label{nu}
\end{figure}

Considering the cylindrical symmetry of the detector it can be assumed
that the maximum allowable step sizes are a function of the helix
axis distance from the origin $r_{c}$ and the absolute value of the helix pitch $|\nu|$:
\begin{equation*}
\text{d}a_{c}^{max} = 
\text{d}a^{max}(r_{c} , |\nu|),
\end{equation*}
\begin{equation*}
\text{d}p_{c}^{max} = 
\text{d}p^{max}(r_{c} , |\nu|),
\end{equation*}
\begin{equation*}
\text{d}\nu^{max} = 
\text{d}\nu^{max}(r_{c} , |\nu|).
\end{equation*}
Here, instead of using $\text{d}x_{c} , \text{d}y_{c}$,
a shift along $\hat{a}$ and $\hat{p}$ is considered as in Figure \ref{ch}.
The resulting maps are illustrated
on Figure \ref{along}, Figure \ref{perp}, and Figure \ref{nu}. They can be directly used to calculate
the total number of iterations in the helix detection algorithm.

The total number of iterations necessary to search for helical tracks in a region $\mathbbm{G}$ of $(r_{c} , |\nu|)$ is:
\begin{equation}
M_{\mathbbm{G}} = 2 \int_{\mathbbm{G}} \frac{2 \pi r_{c}}{\text{d}p^{max}(r_{c} , |\nu|)}
\frac{1}{\text{d}a^{max}(r_{c} , |\nu|) \text{d}\nu^{max}(r_{c} , |\nu|)} \text{d}r_{c} \text{d}|\nu|
\label{in}
\end{equation}
Here
\begin{equation*}
    \frac{2 \pi r_{c}}{\text{d}p^{max}(r_{c} , |\nu|)}
\end{equation*}
is the number of iterations necessary for searching in the whole circle in the $\hat{p}$ direction and
\begin{equation*}
\frac{1}{\text{d}a^{max}(r_{c} , |\nu|) \text{d}\nu^{max}(r_{c} , |\nu|)}
\end{equation*}
is the density of helix parameters in a $\text{d}|\nu|$ by $\text{d}r_{c}$ region. The product of these
two quantities multiplied by $\text{d}r_{c} \text{d}|\nu|$ results in a number of iterations
necessary to investigate a infinitesimal region of $(r_{c} , |\nu|)$. The additional factor of $2$
before the integral is there to account for the helix pitch $\nu = \pm |\nu|$.

For demonstration purposes the region $\mathbbm{G}$ is chosen such that $0.1 \le |\nu| \le 1.1$
and $1.0 \, m \le r_{c}\le 2.0 \, m$. The numerical evaluation of the integral (\ref{in}) results in (PRELIMINARY):
\begin{equation}
M_{\mathbbm{G}} \approx 1.62 \times 10^{11}.
\label{mg}
\end{equation}
In order to find all charged particle trajectories the region $\mathbbm{G}$ should be expanded
making the total number of necessary iterations even larger. This unfortunately indicates
that the algorithm from \cite{Topolnicki_2022} is not a good option for triggering applications.

\section{Summary and conclusions}

The number of iterations necessary in order to carry out the main loop 
of the helix detection algorithm from \cite{Topolnicki_2022} was estimated
using a realistic Monte Carlo simulation of the ATLAS detector. Unfortunetly
this indicates that the numerical complexity of the procedure is to big for
triggering applications. 

The Monte Carlo simulations used provide a good picture of the ATLAS detector. However, the generated
events have a very small number of tracks originating away from the detector. To investigate the effect
of these particles on the step size a larger statistic is needed. Unfortunately it is unlikely that
this would have a significant effect on (\ref{mg}). In addition to the large numerical complexity
choosing helix parameters for the iterations would require constructing a non uniform grid
of helix parameters.

The method proposed in \cite{Topolnicki_2022} indicates not only the existence
or non-existence of a charged particle track in data collected by the detector. 
but also gives estimates of the track's parameters. This copuled with the
algorithm being agnostic to the origin of the track means that it might still
find potential uses in data analysis for hight energy physics experiments. 

\bibliographystyle{unsrt} 
\bibliography{bibl} 

\begin{thebibliography}{1}

\bibitem{Topolnicki_2022}
K.~Topolnicki and T.~Bold.
\newblock Approximate method for helical particle trajectory reconstruction in
  high energy physics experiments.
\newblock {\em Journal of Instrumentation}, 17(08):P08033, aug 2022.

\bibitem{EPJ_WOC_2018}
Mart{\'{\i}}nez~Arantxa Ruiz.
\newblock The {ATLAS} run-2 trigger menu for higher luminosities: Design,
  performance and operational aspects.
\newblock {\em EPJ Web of Conferences}, 182:02083, 2018.

\bibitem{Collaboration_2008}
{ATLAS Collaboration}.
\newblock {The ATLAS Experiment at the CERN Large Hadron Collider}.
\newblock {\em JINST}, 3(08):S08003, aug 2008.

\bibitem{Collaboration2_2017}
{ATLAS Collaboration}.
\newblock Performance of the {ATLAS} trigger system in 2015.
\newblock {\em Eur. Phys. J. C}, 77(5):317, may 2017.

\bibitem{Collaboration1_2017}
{ATLAS Collaboration}.
\newblock {Study of the material of the ATLAS inner detector for Run 2 of the
  LHC}.
\newblock {\em JINST}, 12(12):P12009, dec 2017.

\bibitem{ATLASTDAQAmendment}
Collaboration ATLAS.
\newblock {Technical Design Report for the Phase-II Upgrade of the ATLAS
  Trigger and Data Acquisition System - Event Filter Tracking Amendment}.
\newblock Technical report, CERN, Geneva, Mar 2022.

\bibitem{Bobrovskyi2013}
S.~Bobrovskyi, J.~Hajer, and S.~Rydbeck.
\newblock Long-lived higgsinos as probes of gravitino dark matter at the {LHC}.
\newblock {\em Journal of High Energy Physics}, 2013(2), February 2013.

\bibitem{newbook}
Rudolf Frühwirth and Are Strandlie.
\newblock Pattern recognition, tracking and vertex reconstruction in particle
  detectors.
\newblock chapter~5. Springer Cham, 1988.

\bibitem{ODD}
Corentin Allaire, Paul Gessinger, Julia Hdrinka, Moritz Kiehn, Fabian Kimpel,
  Joana Niermann, Andreas Salzburger, and Stanislava Sevova.
\newblock Opendatadetector, April 2022.

\end{thebibliography}

\end{document}